\documentclass[pdflatex,sn-mathphys-num]{sn-jnl}

\usepackage{graphicx}%
\usepackage{multirow}%
\usepackage{amsmath,amssymb,amsfonts}%
\usepackage{amsthm}%
\usepackage{mathrsfs}%
\usepackage[title]{appendix}%
\usepackage{xcolor}%
\usepackage{textcomp}%
\usepackage{manyfoot}%
\usepackage{booktabs}%
\usepackage{algorithm}%
\usepackage{algorithmicx}%
\usepackage{algpseudocode}%
\usepackage{listings}%

\usepackage{graphicx} 
\usepackage{amsmath, amssymb, amsmath, dsfont, amsthm, graphicx}
\usepackage{tikz}
\usepackage{multirow}
\usetikzlibrary{shapes,decorations,arrows,calc,arrows.meta,fit,positioning,shadings}
\tikzset{
    -Latex,auto,node distance =1 cm and 1 cm,semithick,
    state/.style ={ellipse, draw, minimum width = 0.7 cm},
    point/.style = {circle, draw, inner sep=0.06cm,fill,node contents={}},
    bidirected/.style={Latex-Latex,dashed},
    el/.style = {inner sep=2pt, align=left, sloped}
}

\begin{document}

\title[clinical utility]{Evaluation of clinical utility in emulated clinical trials}
\author*[1]{\fnm{Johannes} \sur{Hruza}}\email{johannes.hruza@sund.ku.dk}
\author[2]{Arvid Sjölander}
\author[3]{Erin Gabriel}
\author[1,4]{Samir Bhatt}
\author[3]{Michael Sachs}

\affil[1]{\orgdiv{Section of Health Data Science and AI}, \orgname{University of Copenhagen}, \orgaddress{\state{Copenhagen}, \country{Denmark}}}

\affil[2]{\orgdiv{Department of Medical Epidemiology and Biostatistics}, \orgname{Karolinska Institutet}, \orgaddress{\state{Stockholm}, \country{Sweden}}}

\affil[3]{\orgdiv{Section of Biostatistics}, \orgname{University of Copenhagen}, \orgaddress{\state{Copenhagen}, \country{Denmark}}}

\affil[4]{\orgdiv{MRC Centre for Global Infectious Disease Analysis}, \orgname{Imperial College London}, \orgaddress{\state{London}, \country{UK}}}

\abstract{Dynamic treatment regimes have been proposed to personalize treatment decisions by utilizing historical patient data, but they may not always improve on the current standard of care. It is thus meaningful to integrate the standard of care into the evaluation of treatment strategies, and previous works have suggested doing so through the concept of \textit{clinical utility}. Here we will focus on the comparative component of clinical utility as the average outcome had the full population received treatment based on the proposed dynamic treatment regime in comparison to the full population receiving the ``standard" treatment assignment mechanism, such as a physician's choice. Clinical trials to evaluate clinical utility are rarely conducted, and thus, previous works have proposed an emulated clinical trial framework using observational data. However, only one simple estimator was previously suggested, and the practical details of how one would conduct this emulated trial were not detailed. Here, we illuminate these details and propose several estimators of clinical utility based on estimators proposed in the dynamic treatment regime literature. We illustrate the considerations and the estimators in a real data example investigating treatment rules for rheumatoid arthritis, where we highlight that in addition to the standard of care, the current medical guidelines should also be compared to any  estimated ``optimal'' decision rule.}



\maketitle



\section{Introduction}
Prognostic risk prediction is often of interest in medical research, with the focus being on high accuracy. Despite the abundance of risk predictions reported in the medical literature, most of them remain unused in clinical settings \cite{Collins2014, Bellou2019}. Sachs et al.\cite{sachs2019aim} pointed out that predicting the future outcomes is not enough, suggesting instead that prognostic modeling research should focus on developing treatment decision rules. By taking potential treatments into account, researchers can more directly address the medical decision-making context. 

Estimating the outcome of using covariates to guide treatment decisions has been explored by \cite{murphy2001} and \cite{Orellana2010}, among others, under the name of dynamic treatment regimes (DTRs). DTRs are rules for making a treatment decision based on a subset of past information. In this framework, one naturally wants to find the optimal treatment regime, which is the regime that maximizes or minimizes the outcome. In general, a DTR can involve time varying treatments but we limit our focus to the case of a treatment decision at a single time point. A simple example of a DTR in the single time point context is the guidelines for rabies post-exposure prophylaxis that assigns treatment according to exposure covariates, such as the nature of the contact with the animal and the risk of rabies transmission \cite{WorldHealthOrganization2018}. For lower-risk exposures (touching or feeding animals or licks on intact skin), no treatment is required. For moderate-risk exposures (minor scratches or abrasions), immediate vaccination is assigned. High-risk exposures (bites, scratches, or contact with broken skin or mucous membranes), require both vaccination and rabies immunoglobulin.

There have been several approaches to estimate the outcome of a dynamic treatment regimes including using Radon–Nikodym derivatives \cite{murphy2001} and inverse probability weighting (IPW) methods  \cite{Tsiatis2019}. The IPW estimators for DTRs can be viewed as an extension of the standard binary IPW estimator \cite{robins1994estimation}. Sachs et al.\cite{sachs2019aim} also pointed out that evaluation of a treatment decision rule should be in comparison to the standard of care (SOC), suggesting that clinical utility should be the target of treatment regime optimization and assessment. This similarly suggests that evaluation of the counterfactual outcome under the ``optimal" DTR alone, is not enough to determine if it should be considered for use in practice.

\textit{Clinical utility} is the average outcome had the full population received treatment based on a regime in comparison to the full population receiving the ``standard" treatment assignment mechanism. This differs from the DTRs paradigm which does not generally include a comparison to the current standard of care. This can potentially be problematic, if, for example, the optimal treatment regime is not superior to the standard of care, and the standard of care is itself too complex to be considered as a regime under the optimization. Physician and patient's choice may be a complex decision process involving information not collected as research data. 

These concepts and practicalities are illustrated in a real-data example in the context of treatment decisions for rheumatoid arthritis (RA). In RA there are many approved second-line treatment options, the choice of which was historically driven by patient and provider preference, in addition to published guidelines from professional associations. Recently there have been updated recommendations published for using clinical data to guide the treatment decision \cite{EULAR2023}. We use a dataset that includes routinely-collected laboratory and clinical data from primary care that was collected before the current guidelines were published. It also spans a period of time when new treatments because available. We use these data and this setting to highlight and discuss different issues concerning what to consider when formulating the standard of care, as well as considering the gap between the current standard of care and any existing treatment guidelines. We also illustrate the practical considerations that exist in such emulated trial settings, such as selecting the time of eligibility, adjusting for confounding, and implementing the guidelines as a concrete DTR. 

In Section 2 we give an overview of notation, preliminary knowledge needed, and summarize the different estimators for clinical utility. We outline the assumptions necessary for the identification of clinical utility and address potential challenges such as positivity violations. In Section 3 we conduct a comprehensive analysis of the finite sample properties of these estimators through simulation studies, shedding light on their performance under varying conditions. In Section 4, we illustrate the estimators and concepts using a real data example related to rheumatoid arthritis.

\section{Preliminaries and Notation}
We will use the following notation throughout the paper. An uppercase letter represents a random variable (e.g., $T$). A realization of a random variable is denoted in lowercase (e.g., $t$). Multivariate random variables and their realizations are in bold (e.g. $\boldsymbol{Z}, \boldsymbol{z}$)

Let $Y$ be the random variable for the outcome of interest, which is assumed to be binary (0/1) or continuous. Let $T$ be the discrete random variable for the observed treatment, with $k \geq 2$ levels. Let $\boldsymbol{Z}$ be a vector of measured covariates of dimension $d$. We will assume throughout that we have $n$ independent identically distributed sampled data points of $(T,\boldsymbol{Z},Y)$. We assume throughout that the observed data satisfies the causal graph represented by the directed acyclic graph below with the associated nonparametric structural equation models (NPSEMs), with independent noise $U_Z, U_T$ and $U_Y$.

\begin{align*}
    \begin{minipage}{0.4\textwidth}
    \centering
    \raisebox{0.4\height}{
    \begin{tikzpicture}[node distance = 0.85 cm and 0.85 cm]
        \node (z) [label = above:\textbf{Z}, point];
        \node (t) [label = below:T, below left = of z, point];
        \node (y) [label = below:Y, point, below right = of z];
        \path (z) edge (t);
        \path (t) edge (y);
        \path (z) edge (y);
    \end{tikzpicture}
    }
\end{minipage}
\hfill
\begin{minipage}{0.4\textwidth}
    \[
    \begin{aligned}
        Z &:= g_Z(U_z) \\
        T &:= g_T(\boldsymbol{Z},U_T) \\
        Y &:= g_Y(\boldsymbol{Z},T,U_Y).
    \end{aligned}
    \]
\end{minipage}
\end{align*}

The NPSEMs framework can be seen as a bridge between graphical and potential-outcome models \cite{Greenland2002}. By defining variables through arbitrary, and flexible, functional forms $(g_Z,g_T,g_Y)$ rather than distributional or counterfactual assumptions alone, this approach explicitly characterizes the variables in the graph as intervenable quantities. This makes consistency — that the observed outcome equals the potential outcome under the observed treatment — an axiom rather than an assumption. Most of the remaining common counterfactual assumptions for identification of an estimand are then implied by the characteristics of the specific NPSEM. Here, mutually independent errors $(U_Z,U_T,U_Y)$ and the lack of any common latent causes implies exchangeability, i.e. no unmeasured confounders. Positivity is still a needed assumption (defined below).

A DTR is a function $f$ that assigns a treatment $t$ based on a covariate vector $\boldsymbol{z}$. The potential outcome $Y(f(\boldsymbol{Z}))$ of an individual is the outcome if the individual is assigned the treatment that the DTR $f$ prescribes for the covariates $\boldsymbol{Z}$. After intervening with a DTR $f$ on the treatment, the equation of $T$ is replaced with $T := f(\boldsymbol{Z})$ while the equations for $\boldsymbol{Z}$ and $Y$ stay the same. The DTR of the observed assignment mechanism is defined by $g_T$, which may be a complex unknown function of both measured and unmeasured covariates. Thus, the expected outcome under the observed treatment assignment mechanism is $E[Y(g_T(\boldsymbol{Z}, U))] = E(Y)$, which  clearly does not fit the provided DAG and NPSEM if $U$ is associated with $Y$, but may fit the DAG if $U=U_T$. There may also be clinical guidelines for patient treatment, which is also a DTR that we will denote as $f_{cgl}$; we will denote the expected values of the outcome enforcing these guidelines as $E[Y(f_{cgl}(\boldsymbol{Z}))]$. 

This allows us to formally define the clinical utility as the difference in expected outcomes comparing two DTRs. In particular, the clinical utility of $f$ as defined in \cite{sachs2019aim} is $E[Y(f(\boldsymbol{Z}))] - E[Y]$, this is in comparison to the observed assignment mechanism. It may be the case that the standard of care has changed over the course of the observation period, and thus the ideal comparison is not to $E[Y]$. One may also wish to compare $f$ to the clinical guidelines $E[Y(f(\boldsymbol{Z}))] - E[Y(f_{cgl}(\boldsymbol{Z}))]$, which is also the clinical utility, as any new decision rule will at best become a new set of guidelines. We discuss these issues further in Section \ref{sec:SOCreasoning}. 

For identification and estimation the clinical utility, we assume \textit{Positivity}, i.e. for any observed $\boldsymbol{z}$ we have that $0<P(T=t\mid \boldsymbol{Z} = \boldsymbol{z})$ for all values of $t$ that we wish to consider, in addition to the NPSEMs given above. Under these assumptions, we can identify $E[Y(f(\boldsymbol{Z}))]$ and estimate it using existing estimators.

\subsection{Estimators} \label{sec: estimators}
Let $f$ be a DTR. Then, under the assumption of positivity and the NPSEM above the estimand of the potential outcome under $f$ can be written as $$
E(Y(f(\boldsymbol{Z})))=E \left[Y \frac{\mathds{1}[f(\boldsymbol{Z})=T]}{P(T=f(\boldsymbol{Z}) \mid \boldsymbol{Z})}\right]. 
 $$ 
 See (\cite{Tsiatis2019} p. 59) or in the Supplementary Materials Section 2 for details. 
This gives rise to a consistent and asymptotically unbiased estimator, given by:
\begin{align}
     \text{ipw}_{nb}&=\frac{1}{n}\sum_{i=1} ^n \frac{y_i\mathds{1}[f(\boldsymbol{z}_i) = t_i]}{\hat{p}(T=t_i\mid \boldsymbol{z}_i)} \label{eq: IPWnb},
\end{align}
where $\hat{p}(T = t_i \mid \boldsymbol{z}_i) = \pi_{nb}(\boldsymbol{z}_i; \hat{\gamma}_{nb})$ is the estimated probability $P(T=t_i \mid \boldsymbol{Z} = \boldsymbol{z}_i)$, commonly called the propensity score, under some model with estimated parameters $\hat{\gamma}_{nb}$. We use the subscript nb, non-binary, in this case because the outcome of the propensity score has as many categories as the treatment. 

Alternatively we can also consider $\mathds{1}[T=f(\boldsymbol{Z})]$ as the binary exposure variable. The estimator with binary (b) outcome of the propensity score is the following: 
\begin{align}
    \text{ipw}_{b}=\frac{1}{n}\sum_{i=1}^n \frac{y_i \mathds{1}[f(\boldsymbol{z}_i) = t_i]}{\hat{p}(\mathds{1}[t_i=f(\boldsymbol{z}_i)]=1 \mid \boldsymbol{z}_i)}, \label{eq: IPWb}
\end{align}
where $\hat{p}(\mathds{1}[t_i=f(\boldsymbol{z}_i)]=1 \mid \boldsymbol{z}_i) = \pi_{b}(\boldsymbol{z}_i; \hat{\gamma}_{b})$ is the estimated probability $P(\mathds{1}[T_i=f(\boldsymbol{z}_i)]=1 \mid \boldsymbol{Z}_i = \boldsymbol{z}_i)$, under some model with estimated parameters $\hat{\gamma}_{b}$.

To obtain consistent estimates, the propensity score model for $\hat{p}(T=t_i\mid \boldsymbol{z}_i)$ has to be correctly specified in \ref{eq: IPWnb} and, similarly, for estimator \ref{eq: IPWb} it is the binary propensity score model for $\hat{p}(\mathds{1}[t_i=f(\boldsymbol{z}_i)]=1 \mid \boldsymbol{z}_i)$ that has to be correctly specified. 

We consider two IPW estimators as both have advantages and disadvantages. An advantage of the first estimator is that it may be easier to use subject matter knowledge to specify $\hat{p}(T\mid \boldsymbol{Z})$, since it is a model for how the observed treatment is assigned. A disadvantage is that when $T$ can take more than 2 levels, it is not a standard binary regression model. The second estimator only ever requires a standard binary regression, but it is for the outcome $\mathds{1}[t_i=f(\boldsymbol{z}_i)]$, which is detached from reality, and thus, expert knowledge is unlikely to help in model specification. Even if $T$ has two levels, the two propensity score models differ because the first model is estimated using the levels of $T$ (e.g., for two levels, $T \in \{0, 1\}$) as the outcome, whereas the second model uses $\mathds{1}[t_i=f(\boldsymbol{z}_i)]$.

It is of note that when using the $\text{ipw}_b$ estimator to compare different DTRs, there could be issues of noncompatibility since the two fitted propensity score models might not be variation-independent. In contrast the $\text{ipw}_{nb}$ requires only one propensity model, even when comparing multiple DTRs. 

Alternatively, we can use g-computation; see supplementary material section 2 for details as well as \cite{Tsiatis2019}. 
The estimator $gc_b$ (g-computation binary) is given by 
\begin{align*}
gc_b=\frac{1}{n}\sum{ \hat{Y}_i(\mathds{1}[t_i=f(\boldsymbol{z}_i)]=1)},
\end{align*}
where $\hat{Y}_i(\mathds{1}[T=f(\boldsymbol{z}_i)]=1)$ is the predicted value from an outcome model $h_b(\boldsymbol{z}_i, t_i; \hat{\alpha}_b)$ for some estimated parameters $\hat{\alpha}_b$. 

Furthermore, $gc_{nb}$ (g-computation non-binary) is given by 
\begin{align*}
gc_{nb} =\frac{1}{n}\sum{ \hat{Y}_i(T=f(\boldsymbol{z}_i))},
\end{align*} 
where $\hat{Y}_i(T=f(\boldsymbol{z}_i))$ is the predicted value from an outcome model $h_{nb}(\boldsymbol{z}_i; \hat{\alpha}_{nb})$ for some estimated parameters $\hat{\alpha}_{nb}$. 

In the first, the outcome model $h_b(\cdot)$, and in the second, $h_{nb}(\cdot)$ for $Y$, must be correctly specified for consistent estimation. For $h_b$,  the predictor variables include $\mathds{1}[T=f(\boldsymbol{z}_i)]$, which we use to directly predict $\hat{Y}_i(\mathds{1}[T=f(\boldsymbol{z}_i)]=1)$ for all subjects. In contrast, in $h_{nb}$, we include $T$ as a predictor variable, and predict $ \hat{Y}_i(T=f(\boldsymbol{z}_i))$ for each subject. We leave the outcome models otherwise unspecified as $h_b(\cdot)$ and $h_{nb}(\cdot)$. 

\section{Simulation study} 
We investigate the finite sample properties of the estimators via simulations. The R code for the simulations can be found in the Supplementary Material and on \url{https://github.com/jhruza/estimate-clinical-utility}. 

The synthetically generated data is based on the DAG above.
We simulate $n \in \{2000,500,200\}$ subjects, where each subject consists of the triplet $(y, t, \boldsymbol{z})$ which is given by:
\begin{align*}
    &y \in \{0,1\} &\text{outcome} \\
    &t \in \{1,2,3\} & \text{received treatment} \\
    &\boldsymbol{z} = \begin{pmatrix}
        z_1 \\
        z_2 
    \end{pmatrix} \in [0,1] \times \{0,1\} & \text{covariates}.
\end{align*}

The covariates $\boldsymbol{Z}=(Z_1, Z_2)$ are sampled from a discrete uniform distribution $Z_1 \sim \mathcal{U}(0,1)$ in discrete steps of $0.2$ and a binomial distribution $Z_2 \sim \mathcal{B}(0.7)$. 
Next, we assign each person a treatment (which will represent the standard of care). This observed treatment allocation differs in Settings 1-3. For a comprehensive description of the data generation process, see Supplementary Material Section 3. Last, we need to simulate the outcome. The generation of the outcome $Y$ relies on the risk function defined by: 
\begin{align} \label{eq:risk_function}
    P(Y=1 \mid \bold{z}, t) = \begin{cases}
    0.5 + 0.5 z_1 - 0.5 z_2 & \text{ if } t=1\\ 
    0.65 - 0.5 z_2 & \text{ if } t=2\\ 
    1 - 0.5 z_1 - 0.5 z_2  & \text{ if } t=3\\ 
    \end{cases}
\end{align}
where $Y=0$ is the desired outcome. 

We explore the properties of the estimators across three distinct settings:

\begin{itemize}
    \item \textbf{Setting 1 (S1)}: In this setting, we choose a DTR to improve the average outcome relative to the standard of care. In fact, it is the optimal DTR. That is, no alternative DTR yields a superior outcome. Here, the clinical utility is positive, reflecting substantial improvement.
    \item \textbf{Setting 2 (S2)}: In this setting, the clinical utility is zero. Despite employing a different DTR, individual treatments are allocated in a manner that produces an identical counterfactual average outcome to that of the standard of care. This scenario presents a nullifying result.
    \item \textbf{Setting 3 (S3)}: Positioned between the extremes of Settings 1 and 2, in this scenario, the clinical utility remains positive. Although the chosen DTR leads to an improved average result, it is less than the optimal regime. Here, we observe a positive but sub-optimal improvement in clinical outcomes.
    \item \textbf{Setting 2 misspecified (S1M)}:Identical to Setting 2 but the models are misspecified.
\end{itemize}
The case where the DTR under investigation performs worse than the SOC is analogous to Setting 1 and 3 but with a change in sign of the clinical utility. As expected, violations of the positivity assumption in these simulations led to instability or undefined values of the IPW estimators, caused by probabilities near zero in the denominator, however the performance of the correctly specified g-computation estimators remained acceptable as long as the extrapolation is not introducing bias.

The following simulation parameters were used: Population size $n \in \{200, 500, 2000\}$, number of simulation iterations $n_{iter} = 2000$, number of bootstraps to construct confidence intervals: $n_{boot} = 500$.  \\

We are comparing the four estimators of the clinical utility of the form $E[Y(f(\boldsymbol{Z}))] - E[Y]$, where $E[Y]$ is given by the mean of $Y$, and $E[Y(f(\boldsymbol{Z}))]$ is estimated using one of the four estimators outlined above;  $\text{ipw}_{nb}$, $\text{ipw}_{b}$, $gc_{nb}$, and $gc_{b}$. 

We specify these as follows in S1-S3:
\begin{enumerate}
    \item $\pi_b$ is a logistic regression, i.e. $\text{logit}(\pi_b) =\boldsymbol{x}^\top \boldsymbol{\gamma}_b $, where $\boldsymbol{x}$ is a design matrix that is a function of $\boldsymbol{z}$ and $\boldsymbol{\gamma}_b$ is the parameter. For this simulation we consider $\boldsymbol{Z}$ to be categorical, we choose our design matrix $\bold{x}$ to contain all interactions. In this case the model is saturated and hence correctly specified. 
    \item $\pi_{nb}$ is a multinomial regression with three levels of $T$, where the  design matrix of the regression contains all interactions of $\boldsymbol{z}$. This is also correctly specified as the model is saturated. The implementation uses the nnet package \cite{nnet}.
    \item $h_b$ is a logistic regression, i.e. $\text{logit}(h_b) =\boldsymbol{x}_b^\top \boldsymbol{\alpha}_b$ where the design matrix $\boldsymbol{x}$ contains all the interactions of $z_i$ and is hence correctly specified.
    \item $h_{nb}$ is a linear regression, i.e. $h_{nb}=\boldsymbol{\alpha}_{nb}^\top \boldsymbol{x}$, where $\boldsymbol{x}$ is a design matrix that encodes the underlying linear risk function \ref{eq:risk_function} and hence is correctly specified. 
\end{enumerate}

In the misspecified setting: 
\begin{enumerate}
    \item $\pi_b$ is a logistic regression where the design matrix encodes $z_1 + z_2$, i.e. the interaction term is missing.
    \item $\pi_{nb}$ is a multinomial regression where the design matrix contains an intercept and $z_1$. 
    \item $h_b$ is a logistic regression where the design matrix contains the intercept and $z_1$.
    \item $h_{nb}$ is a linear regression where the design matrix encodes $0 + \mathds{1}[t=1] + \mathds{1}[t=2] + \mathds{1}[t=3] + z_1 + z_2$.
\end{enumerate}

\subsection{Results}
The results of our simulation study are displayed in Table \ref{fig:result_table}. Notably, for a substantial sample size of $n = 2000$, all estimators across Settings 1 to 3 exhibit minimal bias, coupled with a commendable coverage level hovering around $95\%$. However, as sample sizes decrease, we observe a gradual increase in the standard error. The performance degradation of the two IPW estimators is particularly noteworthy, becoming apparent at the smallest sample size of $n = 200$, accompanied by a notable increase in bias and standard error, which consequently leads to a decrease in coverage accuracy. 
With a sample size of $n=200$, we observe that the variance of the $\text{gc}_{b}$ estimator is larger than that of both IPW estimators, which is surprising given our expectation that the IPW estimators would degrade more quickly in general. In our case this can be explained by the $\text{gc}_b$ model's inconsistency to fit all 44 parameters (due to all the interactions) at such small sample sizes. In contrast, the simpler $\text{ipw}_{n}$ model, which requires only 22 parameters, is reliably fitting all parameters. The $\text{gc}_{nb}$ estimator shows the smallest variance because its correctly specified outcome model is exactly the risk function \ref{eq:risk_function} of the data generating mechanism, which makes it impossible to outperform. 
The estimators $ipw_b$ and $ipw_{nb}$ are almost identical in S1-S3. Since $h_b$ and $h_{nb}$ are the correctly specified models of the propensity score, they are the same for individuals we sum over, i.e. the ones for which $f(z)=t$. In the misspecified setting $ipw_b$ and $ipw_{nb}$ differ as their outcome models for the propensity scores are differently (mis)specified.

\begin{center}
\begin{table*}
     \caption{Simulation results of the clinical utility for the different estimators in settings 1-3 and S2M. The abbreviations are Bias (B), Standard Error (SE) and 95\% confidence interval Coverage (Co).      \label{fig:result_table}}

    \begin{tabular}{|p{0.5cm} p{2cm}||p{1cm} p{1cm} p{1cm}|p{1cm} p{1cm} p{1cm}|p{1cm} p{1cm} p{1cm}| }
     \hline
     & & \multicolumn{3}{c|}{n=2000} &\multicolumn{3}{c}{n=500} &\multicolumn{3}{|c|}{n=200} \\
     &Estimator & B$\times 10^{2}$ & SE$\times 10^{1}$ &Co & B$\times 10^{2}$ & SE$\times 10^{1}$ &Co& B$\times 10^{2}$ & SE$\times 10^{1}$ & Co\\
     \hline
     \multirow{4}{*}{S1}
        & $ipw_b$ & -0.01 & 0.14 & 0.95 & 0.65 & 0.29 & 0.91 & 2.80 & 0.45 & 0.78 \\ 
        & $ipw_{nb}$ & -0.01 & 0.14 & 0.95 & 0.65 & 0.29 & 0.91 & 2.80 & 0.45 & 0.77\\ 
        & $gc_b$ & -0.02 & 0.14 & 0.94 & 0.05 & 0.29 & 0.97 & 0.31 & 0.56 & 0.99 \\ 
        & $gc_{nb}$ & 0.00 & 0.08 & 0.94 & -0.02 & 0.17 & 0.96 & 0.00 & 0.28 & 0.96 \\ 
    \hline
    \multirow{4}{*}{S2}
     & $ipw_b$ & 0.12 & 0.24 & 0.94 & 4.12 & 0.48 & 0.60 & 9.19 & 0.60 & 0.32 \\ 
     & $ipw_{nb}$ & 0.12 & 0.24 & 0.94 & 4.12 & 0.48 & 0.60 & 9.19 & 0.60 & 0.30 \\ 
     & $gc_b$ & -0.06 & 0.23 & 0.96 & 1.47 & 0.57 & 0.97 & 2.59 & 0.87 & 0.99 \\ 
     & $gc_{nb}$ & -0.03 & 0.08 & 0.95 & -0.01 & 0.17 & 0.94 & -0.05 & 0.28 & 0.95 \\ 
     \hline
    \multirow{4}{*}{S3}
     & $ipw_b$ & 0.01 & 0.19 & 0.95 & 2.12 & 0.42 & 0.77 & 6.93 & 0.59 & 0.46 \\ 
     & $ipw_{nb}$ & 0.01 & 0.19 & 0.95 & 2.12 & 0.42 & 0.77 & 6.93 & 0.59 & 0.46 \\ 
     & $gc_{b}$ & -0.02 & 0.19 & 0.95 & 0.45 & 0.43 & 0.96 & 2.00 & 0.77 & 0.99 \\ 
     & $gc_{nb}$ & -0.02 & 0.08 & 0.95 & -0.06 & 0.15 & 0.96 & 0.07 & 0.26 & 0.96 \\ 
     \hline

     \multirow{4}{*}{S2M}
    & $ipw_b$ & -0.08 & 0.40 & 0.97 & 2.15 & 9.19 & 0.97 & 2.24 & 3.43 & 0.83 \\ 
    & $ipw_{nb}$ & 2.62 & 0.24 & 0.79 & 2.69 & 0.52 & 0.86 & 4.67 & 0.78 & 0.71 \\ 
    & $gc_{b}$ & 9.43 & 0.20 & 0.00 & 9.34 & 0.40 & 0.35 & 9.43 & 0.64 & 0.67 \\ 
    & $gc_{nb}$ & 2.07 & 0.11 & 0.51 & 2.05 & 0.21 & 0.84 & 2.08 & 0.34 & 0.90 \\ 
     \hline
    \end{tabular}
   \end{table*}
\end{center}

\section{Rheumatoid arthritis example}  \label{sec: real data example}

We illustrate both the estimators and reasoning suggested above in a real-world data example of patients with rheumatoid arthritis. Rheumatoid arthritis (RA) is a chronic inflammatory disorder that primarily affects the joints but can also cause systemic issues throughout the body \cite{Grassi1998}. Characterized by pain, swelling, and stiffness in the joints, RA can lead to significant disability and decreased quality of life \cite{Matcham2014} if not properly managed. Given its complexity and variability in patient response to treatment, RA presents an interesting scenario for evaluating clinical utility.

The clinical management of RA follows a stepped approach. When individuals are first diagnosed, they will generally receive one of several first-line treatments, which include steroids, hydroxycloraquine, methotrexate, and sulfasalazine. After some time living with the condition, it may worsen despite continuing the initial therapy. At this time, the treatment may be switched or escalated, at what we call the phase II treatment decision. It is at this decision where there is more uncertainty about efficacy. In recent years, there have been many new drugs entering the market, including biologics and JAK inhibitors. Guidelines have been developed and recently updated to help inform this phase II treatment decision \cite{EULAR2023}. The guidelines take into account the individuals' clinical history and some clinical and biochemical prognostic factors. Thus it is of interest to determine the clinical utility of the EULAR 2023 guidelines in comparison to the previous standard of care, and also to determine whether a better DTR can be identified with data-driven methods.  
\subsection{Data} 

The Danish National Laboratory Database includes all results from laboratory tests ordered in-hospital (inpatient and outpatient) from 2015 \cite{arendt2020existing}. The laboratory data have been linked with the Danish National Patient Register and the National Prescribed Drug Register. 

Our dataset from the Danish registers includes information on 1966 patients who meet our eligibility criteria, and records of medications, both in pill form (e.g., sulfasalazine, prednisone, hydroxychloroquine, methotrexate, etanercept, adalimumab) and infusions form (e.g., infliximab, etanercept, adalimumab, golimumab, certolizumab pegol, anakinra, tocilizumab, ustekinumab, secukinumab, abatacept).

The outcome of interest is the log C-reactive protein (CRP) level measured at the earliest time point between 3 months after the treatment switch and up to 1 year after the treatment switch. CRP is a marker of inflammation, hence a smaller outcome is more desirable. Covariates which were thought to possibly influence the outcome and/or the treatment selection included in the models are the phase I treatment, last measured CRP before treatment switch, gender, indicator of being married, indicator of being an immigrant or child of an immigrant, years since RA diagnosis, history of cardiovascular disease, history of chronic obstructive pulmonary disease, and history of osteoperosis. A description of the patients who are included in our emulated target trial is in Table \ref{tab:table1}.

\begin{table}[ht]
\caption{Descriptive statistics of the Danish register data by phase II observed treatment group included in our emulated target trial.}
    \label{tab:table1}
    \centering
    \begin{tabular}{|l|l|l|l|}
    \hline
    & csDMARD (N=1842) & biologics (N=124) & Overall (N=1966) \\
    \hline
    First line treatment & & & \\
    \hline
   hydroxycloraquine & 356 (19.3\%) & 22 (17.7\%) & 378 (19.2\%) \\
    methotrexate & 761 (41.3\%) & 58 (46.8\%) & 819 (41.7\%) \\
    sulfasalazine & 725 (39.4\%) & 44 (35.5\%) & 769 (39.1\%) \\
    \hline
    Pre-switch CRP value (mg/L) & & & \\
    \hline
    Mean (SD) & 13.5 (19.2) & 17.4 (24.1) & 13.7 (19.5) \\
    Median [Q1, Q3] & 7.00 [3.00, 16.0] & 8.90 [3.00, 19.1] & 7.00 [3.00, 16.0] \\
    \hline
    Gender & & & \\
    \hline
   Female & 1352 (73.4\%) & 89 (71.8\%) & 1441 (73.3\%) \\
    Male & 490 (26.6\%) & 35 (28.2\%) & 525 (26.7\%) \\
     \hline
    Married & 1117 (60.6\%) & 72 (58.1\%) & 1189 (60.5\%) \\
    \hline
    Immigrant status & & & \\
    \hline
    Native Danish &  1734 (94.1\%) & 117 (94.4\%) & 1851 (94.2\%) \\
    Immigrant or child of immigrant & 108 (5.9\%) & 7 (5.6\%) & 115 (5.8\%) \\
    \hline
    Duration of RA (years) & & & \\
    \hline
       Mean (SD) & 10.1 (6.28) & 8.65 (6.29) & 10.0 (6.29) \\
    Median [Q1, Q3] & 9.25 [5.06, 14.7] & 6.66 [3.34, 13.4] & 9.14 [4.89, 14.5] \\
  \hline
    Year of RA diagnosis & & & \\
    \hline
      Mean (SD) & 2006 (6.27) & 2008 (6.16) & 2006 (6.28) \\
    Median [Q1, Q3] & 2007 [2002, 2011] & 2010 [2004, 2013] & 2007 [2002, 2012] \\
   \hline
    CVD history & 64 (3.5\%) & 7 (5.6\%) & 71 (3.6\%) \\
    \hline
    Chronic Pulmonary Disease history &  87 (4.7\%) & 5 (4.0\%) & 92 (4.7\%) \\
    \hline
    Oesteoperosis history  &  135 (7.3\%) & <5 & 139 (7.1\%) \\
    \hline
    \end{tabular}
    
\end{table}

\subsection{Reasoning about time zero}
The ideal trial that we would be emulating would have a clearly defined time point where subjects enter the trial and are randomized to either the estimated optimal DTR or the SOC. At the time of entry into the trial, they would also have the necessary biomarkers or other data measured that are inputs to the DTR or SOC regimes. Thus we need to have a similar time zero defined in our emulated trial. The DTR setting differs in an important way from the standard emulated treatment trial because of the requirement of having the necessary biomarkers measured to apply the DTR. If the DTR requires high-dimensional data or a large number of biomarkers, it may be difficult to identify individuals in an observational dataset who have all the required biomarkers measured within a small time window.

This criterion is context-dependent and varies based on the specifics of the setting. In our setting, we are interested in second-line treatment assignment regimes, so we use time zero as the time of first observed treatment switching. Then we look back from that time up to 6 months in the past for a measurement of CRP that is used to define prognostic group. As we have limited our scope to those subjects who switch, this follows how a randomized trial would enroll subjects who have treatment failure and randomize them to either follow the DTR or SOC. Thus, this clearly defines time zero and the window for observation of the necessary biomarkers to apply the DTR. 

In our applied analysis, time zero is clearly defined as the time of treatment switch, since individuals are only eligible for the intervention once. However, in other contexts, such as comparing treatment initiation versus non-initiation, individuals may become eligible multiple times. This complicates the definition of time zero and introduces risks of common biases, such as immortal time bias. One approach to address this challenge is the clone-censor-weight method described by Hernán and Robins \cite{hernan2016using}. For a broader discussion of strategies for handling  time zero, see also \cite{Wang2023}, \cite{MolerZapata2023}.


\subsection{Inclusion and exclusion criteria}
In the ideal trial, people would be eligible if and when they experience first-line treatment failure, are willing and able to take all possible second-line treatments, and for which a treating clinician does not have a specific treatment indication. Failure is determined based on routine monitoring visits to which we do not have access. However, switching to second-line treatment generally only occurs due to failure. In addition, we use the record of biomarker measurements prior to switching as the indication that clinicians and patients are willing to consider multiple treatments, and need further information on which to base that decision. 

To operationalize this ideal trial in our observational data, we include adults who were diagnosed with RA before 1 January 2016 (ICD-10 codes DM05, DM06), and who started first-line treatment with hydroxycloraquine, methotrexate, or sulfasalazine, and who changed or added a biologic or csDMARD during the years 2016, 2017, and 2018. This treatment changing instance would correspond with the phase II treatment decision as described by the EULAR 2023 recommendations \cite{EULAR2023}. We note that although there may be a few patients that would be a part of our eligible population that are not captured based on our operationalization, we are unable to differentiate them from ineligible subjects due to both groups lacking measurements in our data with no differentiating measure. 
A flowchart of the exclusion and inclusion criteria can be found in Section 4 of the Supplementary Materials.

\subsection{Reasoning about standard of care} \label{sec:SOCreasoning}
The standard of care (SOC) is the mean outcome of the population based on the current standard of care or what is sometimes referred to as physicians' choice. Determining the current standard of care, and if it is possible for a DTR that may  become the new clinical guidelines, to actually improve outcomes in relation to the current SOC requires careful consideration. During this consideration, it is important to keep in mind why we wish to make this comparison; we want to make sure in looking for a superior DTR that it improves on the SOC or put another way that, in the absence of guidelines suggesting the DTR, physicians are not already making equal or better treatment decisions. 

If the SOC yields better outcomes than a regime optimized over measured covariates, it implies one of two possibilities: either the estimated optimal is not actually optimal (e.g., due to finite sample bias or model misspecification) or physicians base their decisions on unmeasured prognostic factors. In the latter case, the SOC uses information unavailable to the researcher, which can make it superior to any rule derived solely from the observed data. However, also in this setting the clinical utility is not identified based on $Z$ alone.

Naively, as was suggested in \cite{sachs2019aim}, one can take the mean of the outcome in the current population  
$$\hat{E}[Y]=\frac{1}{n} \sum_{i=1}^n y_i.$$
This is, of course, the observed standard of care, but even with this estimate we need to be mindful that we only include subjects that are within our population of interest, i.e. for which there is positivity, at least theoretically, for receiving the treatment via the regime $f$ that is used in $Y(f(\boldsymbol{z}_i))$, i.e. $0<p(T=t|\boldsymbol{z}_i)$ for all $t \in f(Z)$ and all $i$ in the sample. This may mean to restrict to a particular calendar time range or a particular country where a treatment we wish to consider as part of our regime was available when the data were collected. The solution to this problem depends on the specifics of the setting. 

In this example, we restricted our study to a relatively small period of time during the years 2016 to 2018 when the standard of care was not expected to change very much.  Guidelines for treatment switching existed at the time, though they were different from the EULAR 2023 guidelines. We also restricted our DTR outputs to broad categories of treatment, csDMARDs versus biologics, as the broad categories were available during the entire time period, even though new biologics came on the market.  

It is important to note that the standard of care may be less stable over time than other covariates. As highlighted by \cite{stensrud2024}, the distribution of treatment decisions made by physicians especially when influenced by evolving guidelines can shift in response to new evidence. This instability makes it more difficult to  generalizing clinical utility estimates across time or settings. In contrast, biological covariates such as biomarkers tend to be more stable. Therefore, when evaluating the clinical utility of a decision rule relative to the SOC, it is important to consider the possibility of temporal drift in physician behavior, which may affect the observed outcomes.

\subsection{Reasoning about enforced guidelines}
In many cases, protocols or published guidelines for allocating treatment have been established. Physicians may not follow the guidelines in all cases, and so this may differ from the SOC, even if the guidelines were available during the period the data were generated. Determining if Physicians' implementation of the guidelines of the SOC during a period when the guidelines were being used, and enforcement of guidelines may be of interest. This provides information about how closely the protocol has been followed and, more importantly, if the standard of care improves the outcomes over the guidelines. This gap should always be considered before replacing current guidelines with a new treatment regime. This, of course, is only possible if guidelines were available during the period while the data were generated, which is not the case in our setting. It is, however, also of interest to see if the guidelines, or possibly just time, improved outcomes over the observable SOC. Thus, one should compare both the estimated optimal regime and SOC to enforcing the current clinical guidelines. 

However, as few, if any, datasets will include an indicator that the physician did or did not follow the guidelines, this requires translation of the often vague guidelines to the functions of the observed data. This translation will be contextual and as mentioned above can only be based on observed variables.  

The EULAR guidelines are the guidelines for switching to second line treatment within RA \cite{EULAR2023}. They consist of five overarching principles and ten specific recommendations. To translate these guidelines into a treatment regime $f_{cgl}$, we have chosen to categorize medications into groups that align with those specified in the guidelines. Therefore, we focus on interventions based on treatment groups rather than individual drugs. The guidelines outline five primary treatment groups: TNF inhibitors, interleukin inhibitors, JAK inhibitors, conventional synthetic (sc) DMARDs, and glucocorticoids (GC). For reasons unknown to us, in the Danish register data no individual has been treated with JAK inhibitors (violating the positivity assumption) and hence this treatment group is not allocated in the $f_{cgl}$ even though present in the guidelines. This also implies that our results cannot be directly generalized to settings where JAK inhibitors are available. Some recommendations propose to make a decision based on the present of poor prognostic factors, which is defined through a list of indicators. We have then used these indicators and cross referenced them with the available observations in the Danish register data. With this we can define $f_{cgl}$:

If pre-treatment-switch CRP $<$ 10 mg/L (indicative of poor prognosis factors):  Treat with csDMARD +/- GC (which we label csDMARD). But if pre-treatment-switch CRP $\geq$ 10 mg/L:
Treat with biologics +/- csDMARD (which we label biologics).

\subsection{Estimated optimal regime}
The estimated optimal DTR, denoted by $f_{opt}$, is estimated using a machine learning approach that incorporates random Fourier features (RFF). This method estimates the best treatment decisions based on measured covariates from a random hold-out training sample, ensuring generalizability. The estimated optimal regime should be interpreted as optimal within the space of measured covariates (L-optimal). If unmeasured factors modify treatment effects, the true optimal regime may differ, and the estimand may not be identified. The model has been trained on all available covariates in the Danish register data sample (Table \ref{tab:table1}).

Random Fourier features \cite{NIPS2007_013a006f} are used to approximate complex, nonlinear functions by mapping data into a higher-dimensional feature space, enabling linear models to capture complex relationships between covariates. RFF approximates the kernel function, leading to efficient computation without directly evaluating the kernel. Essentially, the method uses the Fourier perspective of representing functions as sums of sinusoids, which simplifies the computation of kernel-based algorithms.
It reduces computational complexity while still capturing important interactions between covariates. For an introduction, see Milton et al.\cite{Milton2019-kf}.

\subsection{Results}
 
\begin{figure}[h]
    \caption{Estimated clinical utilities with 95\% confidence intervals for the three DTRs.}
    \centering
    \includegraphics[width=\textwidth]{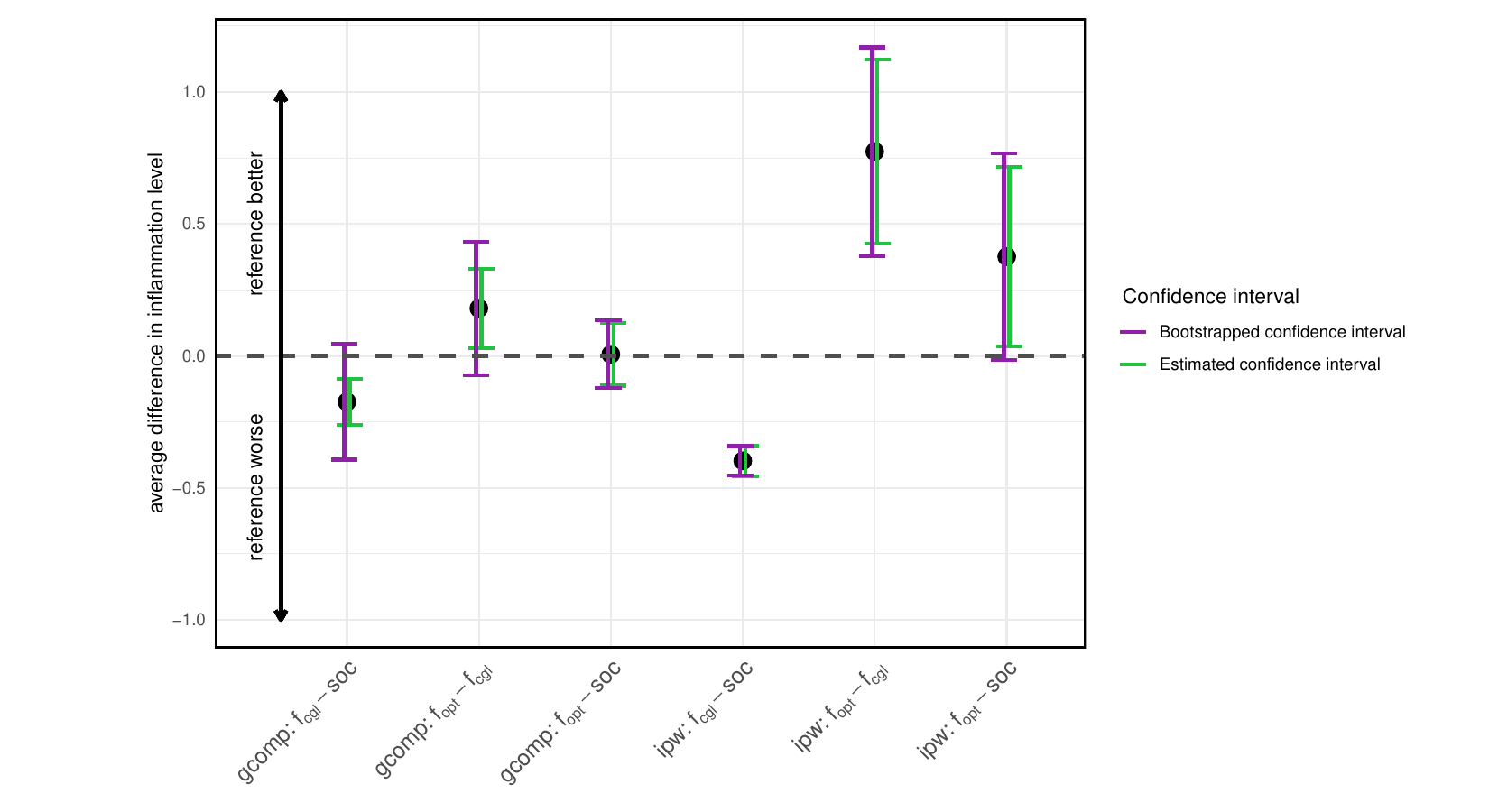}
    \label{fig:plot}
\end{figure}

Figure \ref{fig:plot} shows the clinical utilities with their 95\% confidence intervals, estimated using both the binary g-computation and the binary IPW estimator. We used standard logistic regression implementations for both the outcome and propensity score models. As this is a real world example we cannot guarantee to have observed all confounders. Similarly, the propensity score model and the outcome regression model might be wrongly specified. 
This is the most likely reason for the different confidence intervals between the bootstrapped and the estimated confidence intervals (Supplementary Material Section 2.2) as well as to different estimands between the IPW and g-computation estimators.
The `estimated' confidence intervals are derived analytically using M-estimation theory (sandwich estimator), which relies on asymptotic normality (see Supplementary Material Section 2.1). In contrast, the bootstrapped intervals are obtained by resampling the dataset.

The plot illustrates that simply applying the guidelines compared to the SOC would lead to lower levels of inflammation, indicating the guidelines are an improvement over the standard of care. The estimated clinical utility comparing the estimated optimal regime $f_{opt}$ to the SOC is close to zero, with zero being contained within the confidence intervals. Comparing the estimated optimal regime to enforcing the clinical guidelines we find that the guidelines are never worse and sometimes estimated to be superior, depending on the estimation method. Taken together, there is no evidence to suggest that we would improve patients' outcomes by changing the clinical guidelines to the estimated optimal regime and that if the guidelines are used carefully moving forward, this will improve patient outcomes over the previous SOC, before the guidelines were published. 

This conclusion is not surprising as our estimated optimal regime was derived simply, for demonstrative purposes, and is unlikely to be superior to guidelines put into place by a group of experts. However, it may be somewhat surprising that even a simple estimated optimal regime would do no better than the SOC, which is a important illustration of why the comparison to the SOC is essential for determining if an estimated optimal regime should be considered for use in practice. Interestingly, the guidelines, which are also a function of measured covariates, appear superior to the estimated optimal regime. The estimated optimal regime is derived by optimizing a prediction model (i.e., minimizing a prediction loss), which is distinct from maximizing the clinical utility. Due to this distinction, together with likely finite sample bias and potential model misspecification, the estimated regime is not truly optimal for the target estimand.

As noted above in Subsection \ref{sec:SOCreasoning}, the guidelines are not fully enforceable from the available data, as there will always be room for  physician's input within the treatment decision process. This also illustrates why one might always wish to compare the SOC to the part of the guidelines that are enforceable via the available data, as this speaks to physicians improving on the guidelines. As the guidelines were not published during data accrual, our comparison between the enforceable guidelines and the SOC does not provide such insights. It would, therefore, be of scientific interest to redo this analysis after the guidelines have been in use for a number of years to determine how close enforcing the guidelines is to the observed SOC.

\section{Discussion}
It is crucial to medical decision-making to optimize treatment strategies based on current information and potential interventions. Although this is the target of the DTR methods, less attention has been paid to comparisons of estimated optimal DTR to the SOC.
Sachs et al.\cite{sachs2019aim} suggest the estimand of clinical utility, which compares an estimated optimal decision rule to the SOC. By integrating the standard of care into the evaluation of treatment strategies, it can be easily determined whether it might be worthwhile to attempt implementing novel data-driven DTRs or if the SOC already achieves similar performance.

Via a detailed data example, we have demonstrated the concepts suggested in \cite{sachs2019aim} and extended them while demonstrating how to use them in practice. The suggested estimators are based on familiar DTRs estimators, bridging the gap from single time-point DTR optimization to clinical utility. They provide several additional methods for estimating the estimand above the simple estimator suggested in \cite{sachs2019aim}. The simulation studies demonstrate the finite sample properties of the proposed estimators across various scenarios. Notably, under favorable conditions with a substantial sample size ($n = 2000$), the estimators exhibit minimal bias and nominal confidence interval coverage levels. 

We have outlined several practical details of the sample selection and estimation procedures not highlighted in Sachs et al.\cite{sachs2019aim}. In particular, the enforcement and comparison to the current guidelines were not considered in \cite{sachs2019aim}, which we show in the real-data example to be valuable information. Additionally, we outline the consideration for sample selection for determining the subjects that should be considered withing the eligible population for the standard of care. Finally, in the real data example we demonstrate how one can select a time zero for the emulated trial comparing the estimated optimal decision rule to the SOC or enforced guidelines. 

As we have seen with the data example, if the SOC differs across populations or evolves over time, the clinical utility may also differ as it is relative to the SOC. Conditions that allow generalization to different target distributions may be possible to derive by adapting methods from the transportability literature \cite{bareinboim2013general}. For example, if we assume the target population has the same distributions of treatment options and covariates then we can likely apply our results to this population as well. Instead of attempting to transport the clinical utility, updating the clinical utility of the estimated optimal DTR over time or calculating it in a new setting where the SOC may differ is advisable. Additionally, as with all prediction or machine learning models, data drift has to be considered from the training population in comparison to the target population \cite{Sahiner2023}, as this can cause performance deterioration. 

This suggests that not only should the clinical utility of an optimal DTR be reevaluated with the changing SOC, but that the DTR should potentially be rederived in new settings and as patients' characteristics shift within the same population.  Since the clinical utility is dependent on the specific SOC of the analyzed population, we recommend explicitly characterizing the observed treatment patterns (such as treatment distributions within subgroups) to aid any future meta-analysis or comparisons.

Our analysis focused on a single treatment decision for clarity and feasibility. Many clinical contexts, however, involve multiple sequential decisions, and extending our framework to such settings would be of great interest. Multi stage regimes introduce additional complexity, including the need to model time varying confounding, ensure sequential exchangeability, and maintain positivity across treatment histories. These challenges are well documented in the literature on dynamic treatment regimes \cite{Tsiatis2019}. Furthermore, comparing sustained regimes to the standard of care is complicated by the fact that observed post-baseline treatments may not correspond to the natural course under earlier interventions \cite{Young2014}. Future work should explore these extensions to better capture the longitudinal nature of treatment decisions.

We see comparing any estimated optimal decision rule to the SOC and the current guidelines as an alternative method to including the physician's decision in the optimization, as suggested by Stensrud et al.\cite{stensrud2024}, or other related works, such as  Levis et al.\cite{levis2024intervention}, which propose a measure of the potential benefit of a targeted intervention compared to status quo, conditional on covariates, which is closely related to clinical utility. We note that the \citet{levis2024intervention} approach also allows for the incorporation of supply constraints, making it particularly useful in settings where the intervention of interest is expensive or difficult to obtain. Such restrictions could also be included in the derivation of any DTR, prior to evaluation against SOC using clinical utility. 

Our hope is that this work both understandably provides the methods and demonstrates their use to underscore the importance of integrating the standard of care into treatment optimization processes to enhance clinical decision-making and ultimately improve patient outcomes. Future research exploring alternative methodologies and other real world applications of the proposed estimators may further advance our understanding and application of clinical utility in medical practice.

\subsection*{Funding}
SB acknowledges funding from the MRC Centre for Global Infectious Disease Analysis (reference MR/X020258/1), funded by the UK Medical Research Council (MRC). This UK funded award is carried out in the frame of the Global Health EDCTP3 Joint Undertaking. SB is funded by the National Institute for Health and Care Research (NIHR) Health Protection Research Unit in Modelling and Health Economics, a partnership between UK Health Security Agency, Imperial College London and LSHTM (grant code NIHR200908). Disclaimer: “The views expressed are those of the author(s) and not necessarily those of the NIHR, UK Health Security Agency or the Department of Health and Social Care.”. SB acknowledges support from the Novo Nordisk Foundation via The Novo Nordisk Young Investigator Award (NNF20OC0059309) which also supports JH.  SB acknowledges the Danish National Research Foundation (DNRF160) through the chair grant.  SB acknowledges support from The Eric and Wendy Schmidt Fund For Strategic Innovation via the Schmidt Polymath Award (G-22-63345). EEG and MCS are funded in part by NNF Grant NNF22OC0076595 and the Pioneer Centre SMARTbiomed.


\end{document}